\documentstyle[12pt,titlepage]{article}
\input epsfig.sty

\font\grande=cmr9.5 scaled \magstep4
\font\medio=cmr9.5 scaled \magstep2
\outer\def\beginsection#1\par{\medbreak\bigskip
      \message{#1}\leftline{\bf#1}\nobreak\medskip
\vskip-\parskip
      \noindent}

\def\laq{\raise 0.4ex\hbox{$<$}\kern -0.8em\lower 0.62
ex\hbox{$\sim$}}
\def\gaq{\raise 0.4ex\hbox{$>$}\kern -0.7em\lower 0.62
ex\hbox{$\sim$}}

\begin{document}
\bibliographystyle {unsrt}

\titlepage

\begin{flushright}
CERN-PH-TH/2006-150
\end{flushright}

\vspace{15mm}
\begin{center}
{\grande Non-topological gravitating defects}\\
\vspace{5mm}
{\grande in five-dimensional anti-de Sitter space}\\
\vspace{15mm}
 Massimo Giovannini 
 \footnote{Electronic address: massimo.giovannini@cern.ch} \\
\vspace{6mm}

\vspace{0.3cm}
{{\sl Centro ``Enrico Fermi", Compendio del Viminale, Via 
Panisperna 89/A, 00184 Rome, Italy}}\\
\vspace{0.3cm}
{{\sl Department of Physics, Theory Division, CERN, 1211 Geneva 23, Switzerland}}
\vspace*{2cm}

\end{center}

\vskip 2cm
\centerline{\medio  Abstract}
A class of five-dimensional warped solutions is presented. The geometry 
is everywhere regular and tends to five-dimensional anti-de Sitter space for large absolute values of the bulk coordinate. The physical features of the solutions 
change depending on the value of an integer parameter.  
In particular, a set of solutions describes generalized gravitating kinks where 
the scalar field interpolates between two different minima of the potential. 
The other category of solutions describes  instead
gravitating defects where the scalar  profile is always finite and reaches the same constant asymptote both for positive and negative values of the bulk coordinate. In this sense the profiles are non-topological. The physical features of the zero modes are  discussed.
\noindent

\vspace{5mm}

\vfill
\newpage
In the presence of infinite extra-dimensions \cite{f1} (see also \cite{f1a,f1b}) 
fields of various spin are localized around  higher dimensional 
gravitating defects whose properties determine, at least partially,
the features of the localized interactions. 
Consider, therefore, one of the simplest incarnations of this idea 
and suppose that there is only one infinite extra dimension that will be denoted, in 
what follows, by $w$. The five-dimensional line element can then be written as 
\begin{equation}
ds^2 = g_{ A B} d\,x^{A} d\, x^{B} = a^2(w) [ \eta_{\mu\nu} d\,x^{\mu} d\,x^{\nu} - dw^2],
\label{metric}
\end{equation}
where $\eta_{\mu\nu}$ is the Minkowski metric with signature mostly minus; 
 the Latin (uppercase) indices run over all the five dimensions while the Greek 
 indices run over the $(3+1)$ observable dimensions. The coordinate $w$ 
 runs continuously from $-\infty$ to $+\infty$.
In the situation described by Eq. (\ref{metric}), five-dimensional domain-wall 
solutions are known to exist \cite{f2,f3,f4,f5} (see also \cite{f5a,f5b,f5c}) 
and they have the structure of gravitating kinks whose associated geometry 
is rather similar to the one generated by 3-brane sources supplemented by 
a negative cosmological constant \cite{f5d}.
Five-dimensional gravitating kinks can arise both in the case 
of Einstein-Hilbert gravity and in the case of quadratic gravity theories 
of Euler-Gauss-Bonnet type (see, for instance, \cite{f6,f6a1,f6a2,f6a3} 
and references therein). 

The scalar-tensor action adopted for illustrating the present considerations will then be given by 
\begin{equation}
S = \int d^{5} x \sqrt{|g|} \biggl[ - \frac{R}{2 \kappa} + \frac{1}{2 } 
g^{A B}\partial_{A} \phi \partial_{B} \phi  - U(\phi)\biggr],\qquad \kappa = 
8\pi G_{5} = 8\pi M_{5}^{-3},
\label{action}
\end{equation}
leading to the equations \footnote{In the following, the prime will always denote a derivation 
with respect to the bulk coordinate $w$.}
\begin{eqnarray}
&& {\mathcal F}^2 = \frac{\kappa}{6} \biggl[ \frac{{\phi'}^2}{2} - U a^2\biggr],
\qquad {\mathcal F}' = - \frac{\kappa}{4} \biggl[ {\phi'}^2 + \frac{2}{3} U a^2\biggr],
\label{I}\\
&& \phi'' + 3 {\mathcal F} \phi' - \frac{\partial U}{\partial\phi} a^2 =0,
\label{II}
\end{eqnarray}
where ${\mathcal F} = a'/a$.
A consistent solution of Eqs. (\ref{I}) and (\ref{II}) can be obtained 
in the form 
\begin{eqnarray}
&& a(w) = [ (b w)^{2 \nu} + 1]^{- \frac{1}{2\nu}},\qquad \nu \geq 1,
\label{aw}\\
&& \phi(w) = v \pm \frac{1}{\beta} \,\arctan[(b w)^{\nu}],
\label{phi}\\
&& U(\phi) = U_{0}\, \Lambda(\phi)^{\frac{\nu -1}{\nu}} [ ( 2 \nu -1) - ( 3 + 2 \nu) \Lambda(\phi)]
\label{Uphi}
\end{eqnarray}
where $b$ is a parameter (with dimensions of inverse length) related with the thickness of the scalar profile and where:
\begin{equation}
\Lambda(\phi) = \sin^2{(\phi - v)},\qquad \beta = \sqrt{\frac{\kappa}{3} \biggl(\frac{\nu^2}{2\nu -1}\biggr)},
\qquad U_{0} = \frac{3 b^2 }{2\kappa}.
\label{DEF}
\end{equation}
In Eq. (\ref{phi}), 
$v$ arises as an integration constant with the same dimensions of $\beta^{-1}$. 
In Eqs. (\ref{aw}), (\ref{phi}) and (\ref{Uphi}) $\nu$ is a positive 
integer (i.e. $\nu \geq 1$). 
Since the bulk coordinate $w$ may take both positive and negative 
values, if $\nu$ would be rational or even real,  the functions defining the solution 
will may become imaginary \footnote{From a swift inspection of Eqs. (\ref{aw}), 
(\ref{phi}) and (\ref{Uphi}) it may seem that the solution can be continued 
also for negative values of $\nu$. This is not correct since, if $\nu <0$, ${\phi'}^2$ 
becomes negative, or, equivalently, the parameter $\beta$ defined in Eq.  (\ref{DEF})
becomes imaginary.}.
The curvature invariants pertaining to the solution defined by Eqs. (\ref{aw}), (\ref{phi}) and (\ref{Uphi}) can be simply computed and they are 
\begin{eqnarray}
&& R^2 = 16 \, b^4 \frac{ (b w)^{ 4 ( \nu -1)} [ (2 - 4 \nu) + 5 (b w)^{2 \nu}]^2}{
[ 1 + (b w)^{2\nu}]^{4 - \frac{2}{\nu}}}, 
\label{Rsqex}\\
&& R_{A B} R^{A B} = 4\, b^4 \frac{ (b w)^{2 ( \nu - 2)}[ 20 (b w)^{6 \nu} + 5 ( 1 - 2\nu)^2 
(b w)^{ 2\nu} + 16 ( 1 - 2\nu) (b w)^{4 \nu}]}{[ 1 + (b w)^{2\nu}]^{4 - \frac{2}{\nu}}},
\label{Riccisqex}\\
&& R_{A B C D} R^{A B C D} = 8 b^4 \frac{ (b w)^{4 (\nu - 1)} [ 5 (b w)^{ 4 \nu} + 2 ( 1 - 2\nu)^2 + 
4 (b w)^{2 \nu} ( 1 - 2 \nu)]}{[ 1 + (b w)^{2\nu}]^{4 - \frac{2}{\nu}}},
\label{Riemannsqex}
\end{eqnarray}
where $R_{A B C D}$, $R_{ A B}$ and $R$ are, respectively, 
the Riemann tensor, the Ricci tensor and the Ricci scalar. In the case of the metric (\ref{metric}), the Weyl invariant vanishes. 
Since $\nu \geq 1$,  Eq. (\ref{aw}) implies that  $a(w) \to |b w|^{-1}$ in 
the limit $|b w| \to \infty$, i.e. for values of the bulk coordinate 
much larger than the thickness of the configuration. Since $\nu \geq 1$ is a positive 
integer, the curvature invariants do not have poles for any finite value of $w$. The quantity $w_{0} = b^{-1}$ 
is the radius of the (asymptotic) ${\mathrm AdS_{5}}$ space. Consistently with this behaviour, the explicit form of the curvature invariants 
goes to a constant for $|bw|\to \infty$. 

If $\nu$ is {\em odd}, i.e. $\nu = 2 \, m \, + \, 1$ with $m = 0,\, 1,\,2,\,3,\,...$, 
Eq. (\ref{phi}) implies that  $\beta (\phi - v)$ varies between $-\pi/2$ and $\pi/2$. 
 The plus sign in Eq. (\ref{phi}) corresponds to the kink solution while the 
minus sign corresponds to the anti-kink solution. 
In the case of one spatial dimension, spatial infinity consists of two points, i.e. 
$\pm \infty$; a topological charge is then customarily defined for the characterization  
of  $(1+1)$-dimensional defects such as the ones 
arising in the case of sine-Gordon system \cite{f7}. 
When $\nu$ is odd, therefore, we will have that the topological charge 
does not vanish and is given, in particular, by 
\begin{equation}
Q_{m} = \frac{1}{2\pi} \int_{-\infty}^{\infty} d w \frac{\partial \phi}{\partial w} = \pm \frac{\sqrt{4 m + 1}}{ 2 m + 1} \lambda, \qquad \lambda = \sqrt{\frac{3}{4\kappa}}, 
\label{TC}
\end{equation} 
where the plus and the minus signs correspond, respectively, to a kink and to an anti-kink solution.

If $\nu$ is {\em even}, i.e. $\nu = 2 n$ with $n= 1,\,2,\,3,\,...$, 
from Eq. (\ref{phi}), $\beta(\phi - v)$ goes, asymptotically, to the same value both for $w\to -\infty$ and for $w\to +\infty$. Therefore, applying the definition reported in the first equality of Eq. (\ref{TC}), we will 
have, in this case that $Q_{n}=0$ in spite of the sign appearing in Eq. (\ref{phi}). 
This second class of solutions
seems then to describe more non-topological rather than topological defects.  It should be clear that five-dimensional gravity is essential in order to have this type of profiles. In the absence of gravity, non-topological defects in 
$(1+1)$ dimensions are connected with an additive conservation law, so that one should demand that the system 
contains, at least, a complex scalar field with global $U(1)$ symmetry \cite{f7} or, equivalently, two real scalar fields \cite{f8}. 
Here, however, because of the presence of gravity, bell-like profiles can arise even if  $\phi$  is not complex as the solution (\ref{phi}) demonstrates explicitly when $\nu$ is even.

For graphical illustration, it is 
practical to rescale $\phi$ through $\beta$ in such a way that the 
rescaled field, i.e.  $\overline{\phi}(w) = \beta \phi (w)$ is dimensionless.  
In the following, when not otherwise stated, we will also fix, without loss
of generality, $v =0$.
\begin{figure}
\begin{center}
\begin{tabular}{|c|c|}
      \hline
      \hbox{\epsfxsize = 7.6 cm  \epsffile{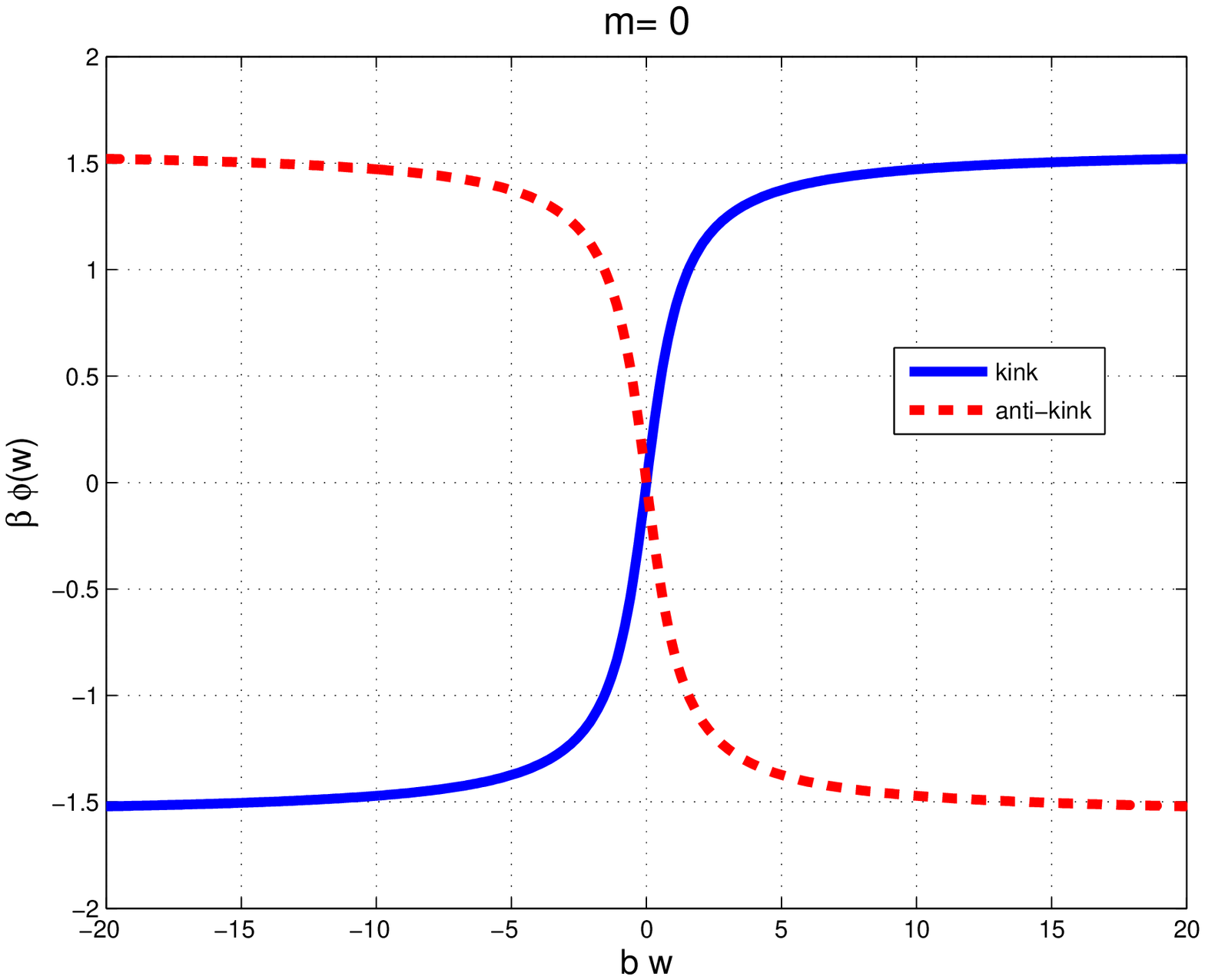}} &
      \hbox{\epsfxsize = 7.6 cm  \epsffile{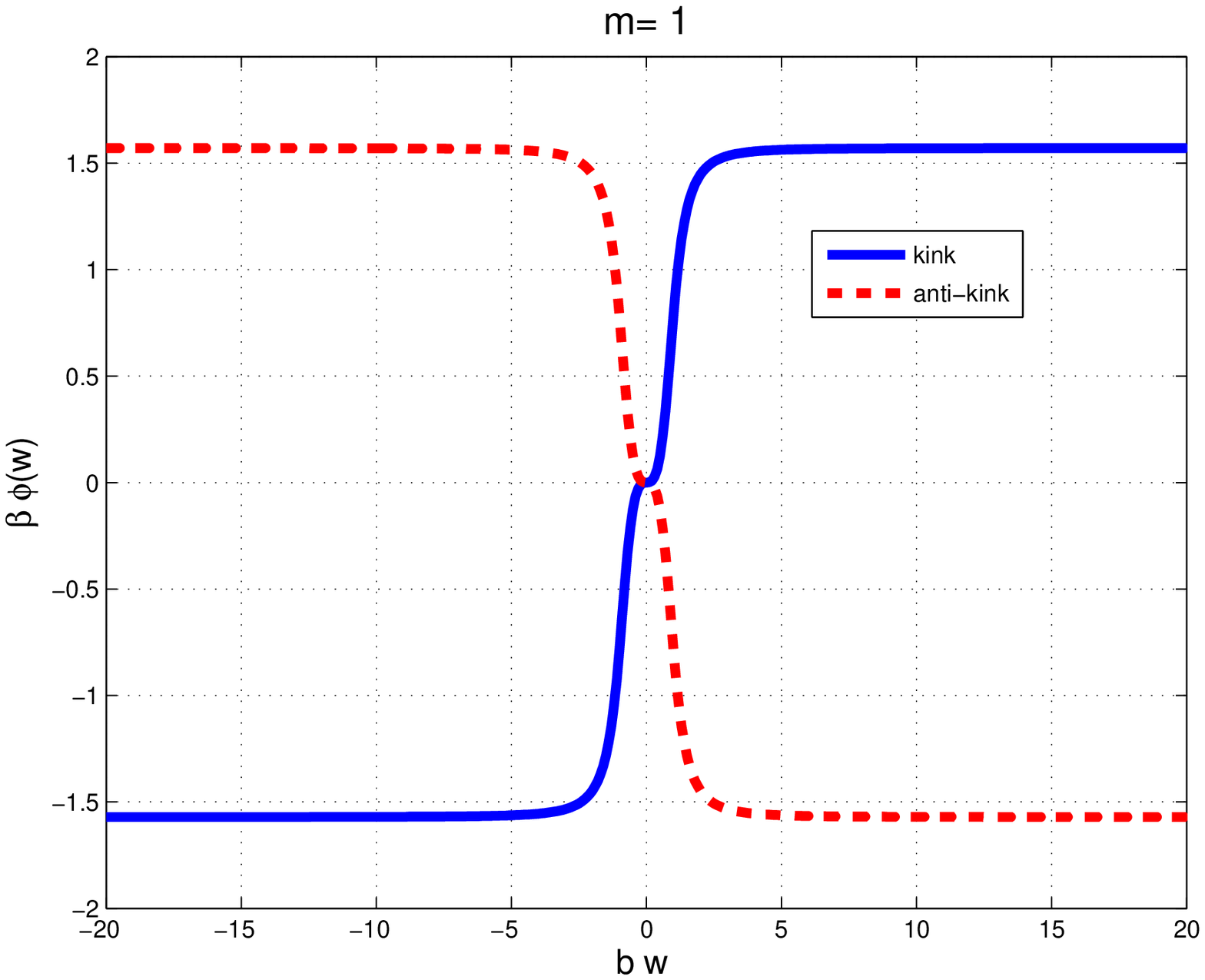}}\\
      \hline
\end{tabular}
\end{center}
\caption[a]{Two cases of kink and anti-kink solutions 
are illustrated. They both arise for odd $\nu$ and, in particular, 
for $\nu =1$ (plot at the left) and $\nu =3$ (plot at the right). Recall that, when $\nu$ is odd, it is conventionally parametrized as  $\nu = 2 m + 1$.}
\label{F1}
\end{figure}
The cases $m =0$ and $m=1$ (i.e. $\nu =1$ and $\nu = 3$) 
are illustrated, respectively, in the left and in the right plot of Fig. \ref{F1}. 
As $m$ increases an intermediate 
plateau develops close to $w =0$ (see the right plot in Fig. \ref{F1} and the left 
plot in Fig. \ref{F2}).
\begin{figure}
\begin{center}
\begin{tabular}{|c|c|}
      \hline
      \hbox{\epsfxsize = 7.6 cm  \epsffile{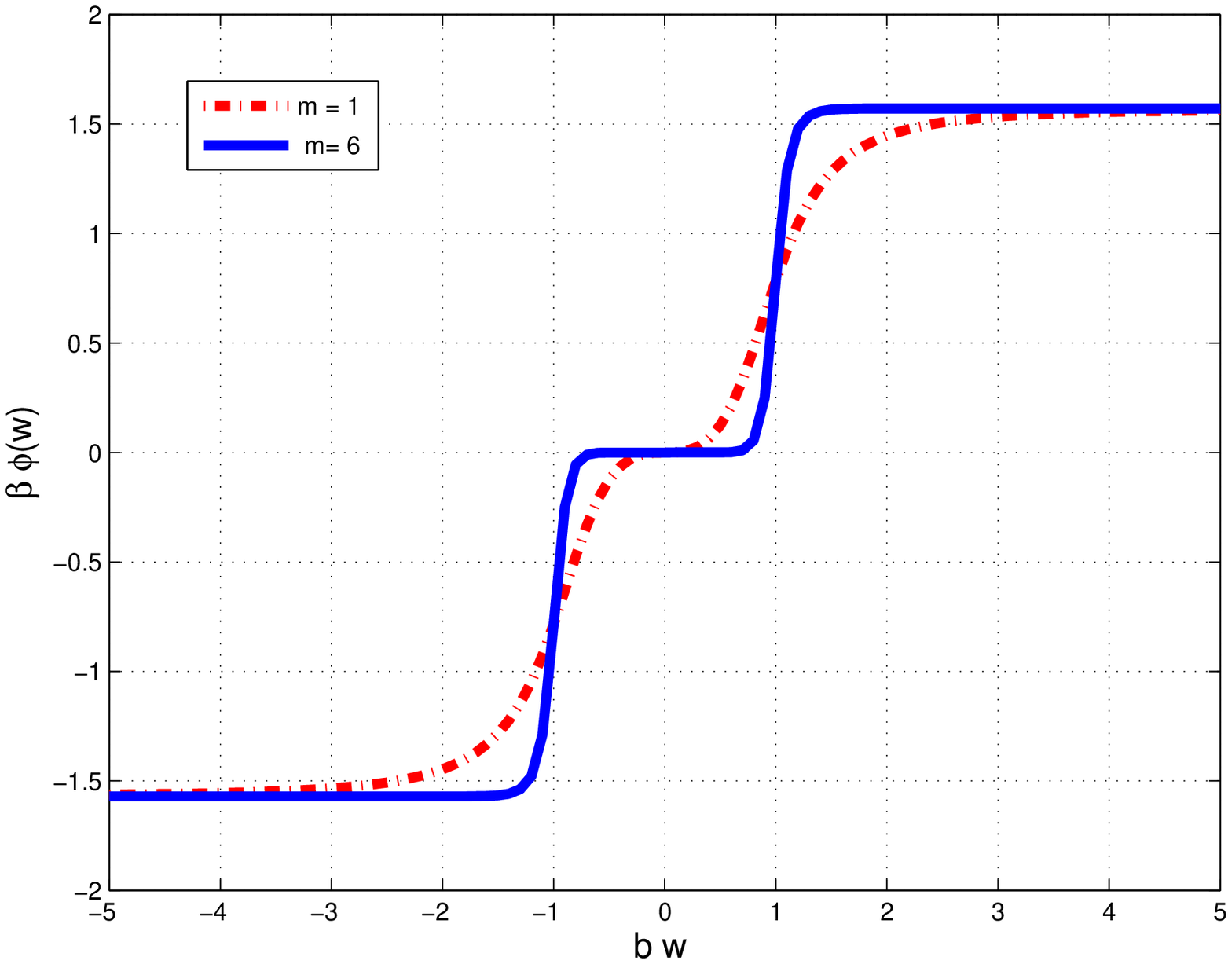}} &
      \hbox{\epsfxsize = 7.6 cm  \epsffile{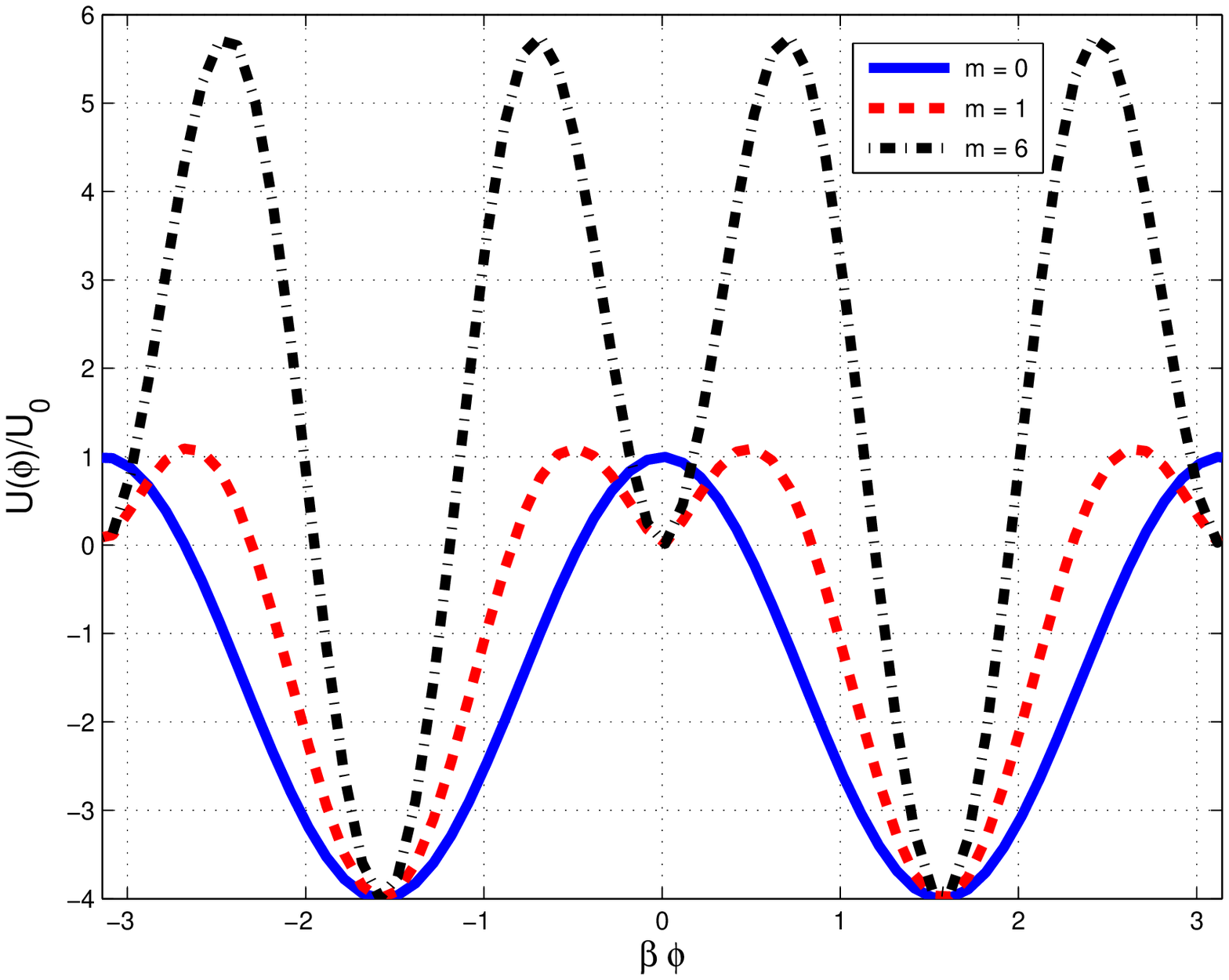}}\\
      \hline
\end{tabular}
\end{center}
\caption[a]{In the left plot 
the behaviour of $\phi$ is illustrated for large (odd) $\nu$ (as in Fig. \ref{F1}, $\nu = 2 m +1$). In the plot at the right the potential $U(\phi)$ is 
reported for three different values of $m$ as specified in the legend.}
\label{F2}
\end{figure}
In the case $m=0$ (full line in the right plot of Fig. \ref{F2}) the potential 
is of sine-Gordon type and it is, according to Eq. (\ref{Uphi}), 
$U(\phi) = U_{0}( 5 \cos{2 \beta\phi} -3)/2$.
The minima of the potential are located, for $\nu =1$,
in $-\pi/2$ and in $+\pi/2$ (see right plot of Fig. \ref{F2}).
By looking simultaneously at Fig. \ref{F1} (full line in the 
left plot) and at Fig. \ref{F2} (full line in the right plot) it appears that the kink 
solution connects the minimum in $-\pi/2$ to the the minimum 
in $\pi/2$ and $\beta \phi$ correctly interpolates between these two values.
This situation reminds a bit the sine-Gordon system in $(1+1)$ dimensions 
\cite{f7} where, however, the potential vanishes at the minima while here it is 
negative due to the gravitating nature of the solution.
As $m$ increases 
the potential develops, at the centre of the interval  of periodicity, 
a second (local) minimum which is located, for the interval chosen 
in Fig. \ref{F2}, in $\phi =0$. Since the minimum is only local (and not 
global) the field does not settle down and finally reaches the true 
global minimum in $\pi/2$.  As $m$ increases further (dot-dashed 
lines in both plots of Fig. \ref{F2}), the 
local minimum becomes more and more pronounced and the 
length of the intermediate plateau in $\phi$ gets larger (see Fig. \ref{F2}, left 
plot).  According to Eq. (\ref{phi}),
both $\phi'$ and ${\phi'}^2$ are always finite and regular for every 
value of the bulk radius.  In Figs. \ref{F1} and \ref{F2} $\beta v =0$ has been 
assumed. 

\begin{figure}
\begin{center}
\begin{tabular}{|c|c|}
      \hline
      \hbox{\epsfxsize = 7.3 cm  \epsffile{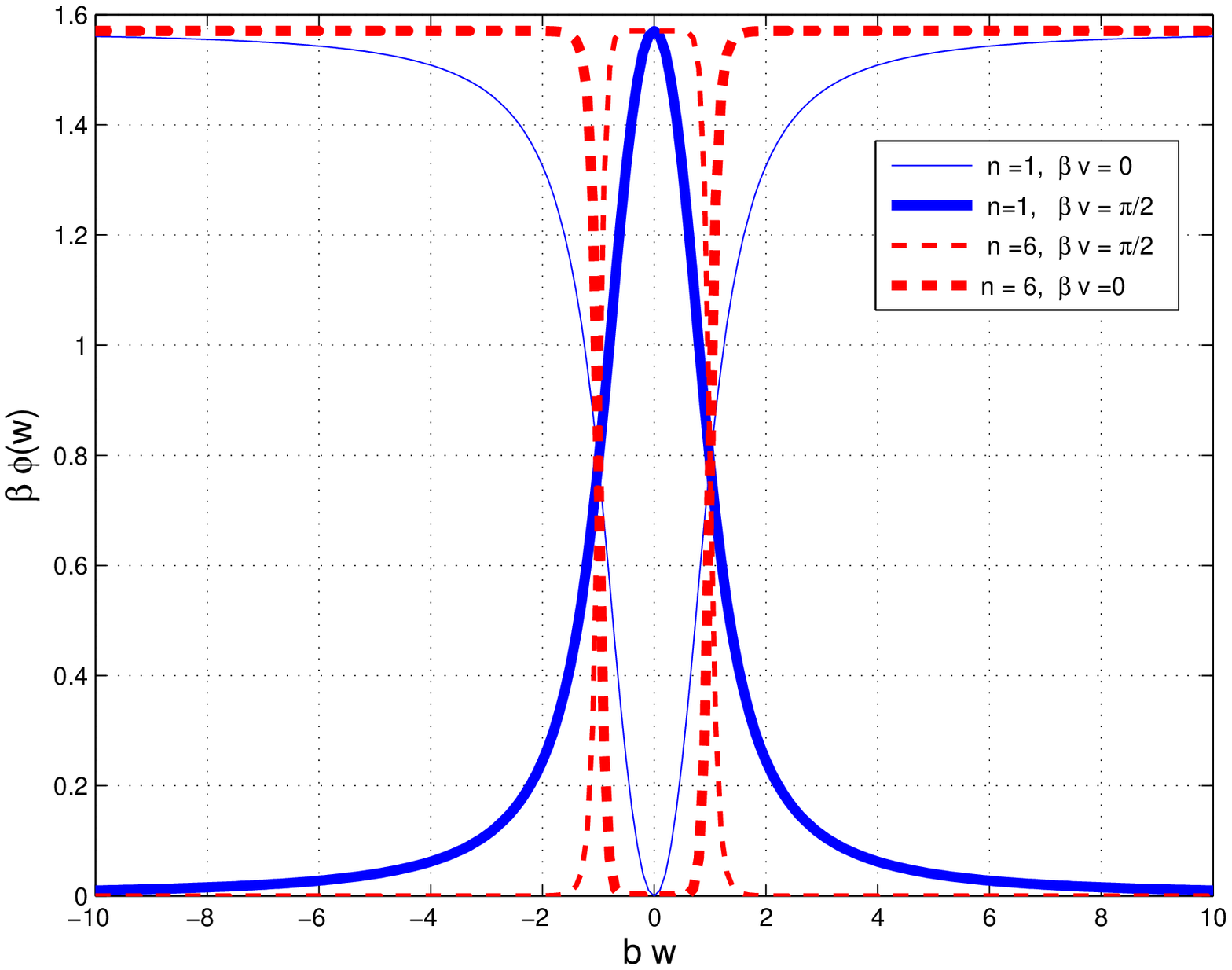}} &
      \hbox{\epsfxsize = 7.6 cm  \epsffile{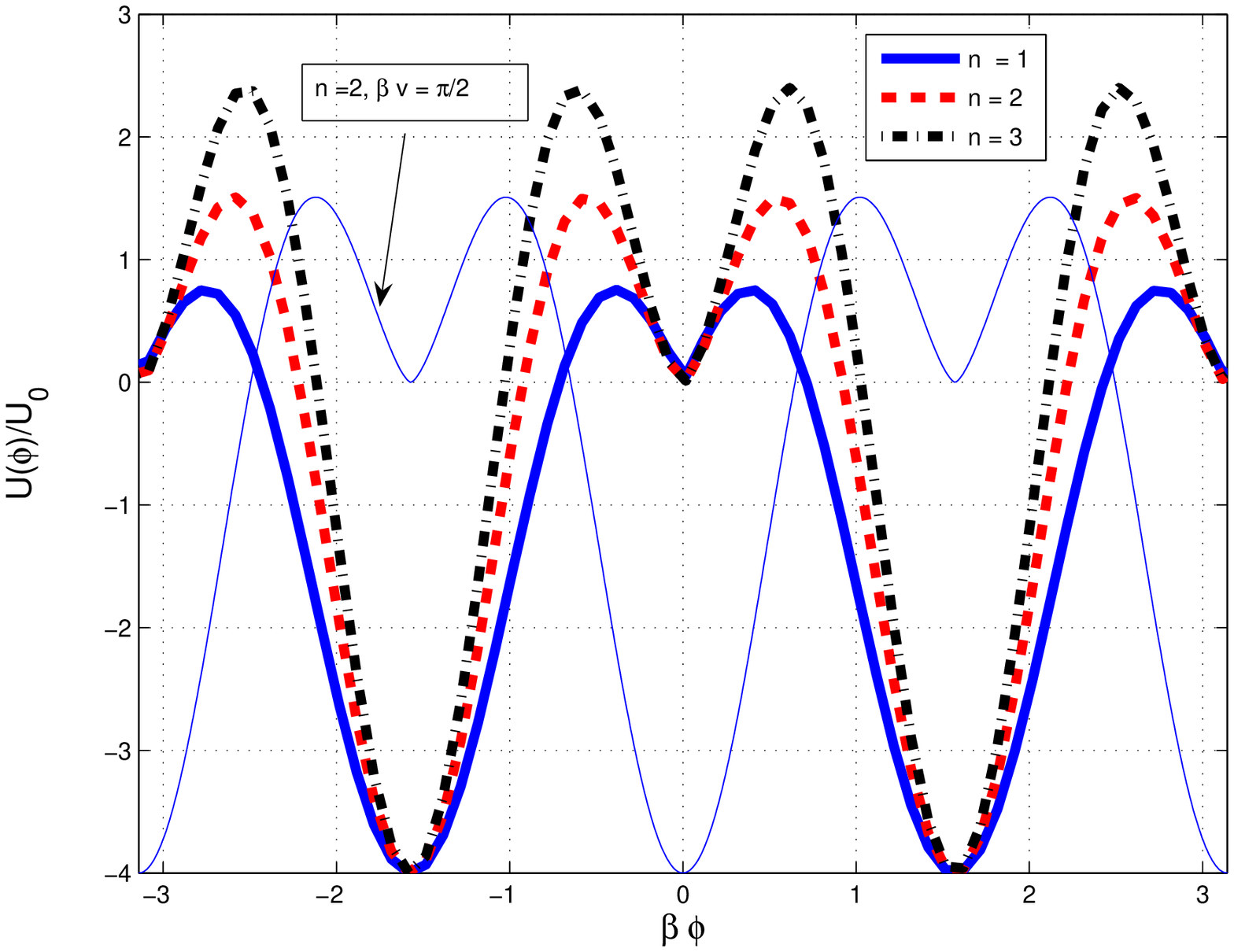}}\\
      \hline
\end{tabular}
\end{center}
\caption[a]{The profiles of the scalar field (plot at the left) and the potential (plot at the 
right) are illustrated for even values of $\nu$. In this case we parametrize 
$\nu = 2 n$ with $n$ positive integer. The cases $n=1$ and $n=6$ are reported (plot at the left), respectively, with the full and thin lines. The thick lines (plot at the left) refer 
to different boundary conditions as specified in the legend. }
\label{F3}
\end{figure}

If  $\nu$ is even the scalar profile goes, asymptotically, to the same value 
for $w \to \pm \infty$.
In Fig. \ref{F3}
we report the profile of $\beta \phi$ and its related potential
for few values of even $\nu$ and for two different values of $\beta v$ (i.e.
$\beta v =0$ and $\beta v = \pi/2$).
In the left plot of Fig. \ref{F3} the scalar field is illustrated as a function of the 
bulk radius for two different values of $\beta v$ 
(i.e. $\beta v=0$ and $\beta v = \pi/2$). 
 By increasing the value of $n$ the width of $\beta\phi$ 
 increases (dashed line in the left plot of Fig. \ref{F3}). 
 Given the properties of this second class of solutions 
 the case of even $\nu$ resembles the one of a non-topological defect. 
 
\begin{figure}
\begin{center}
\begin{tabular}{|c|c|}
      \hline
      \hbox{\epsfxsize = 7.6 cm  \epsffile{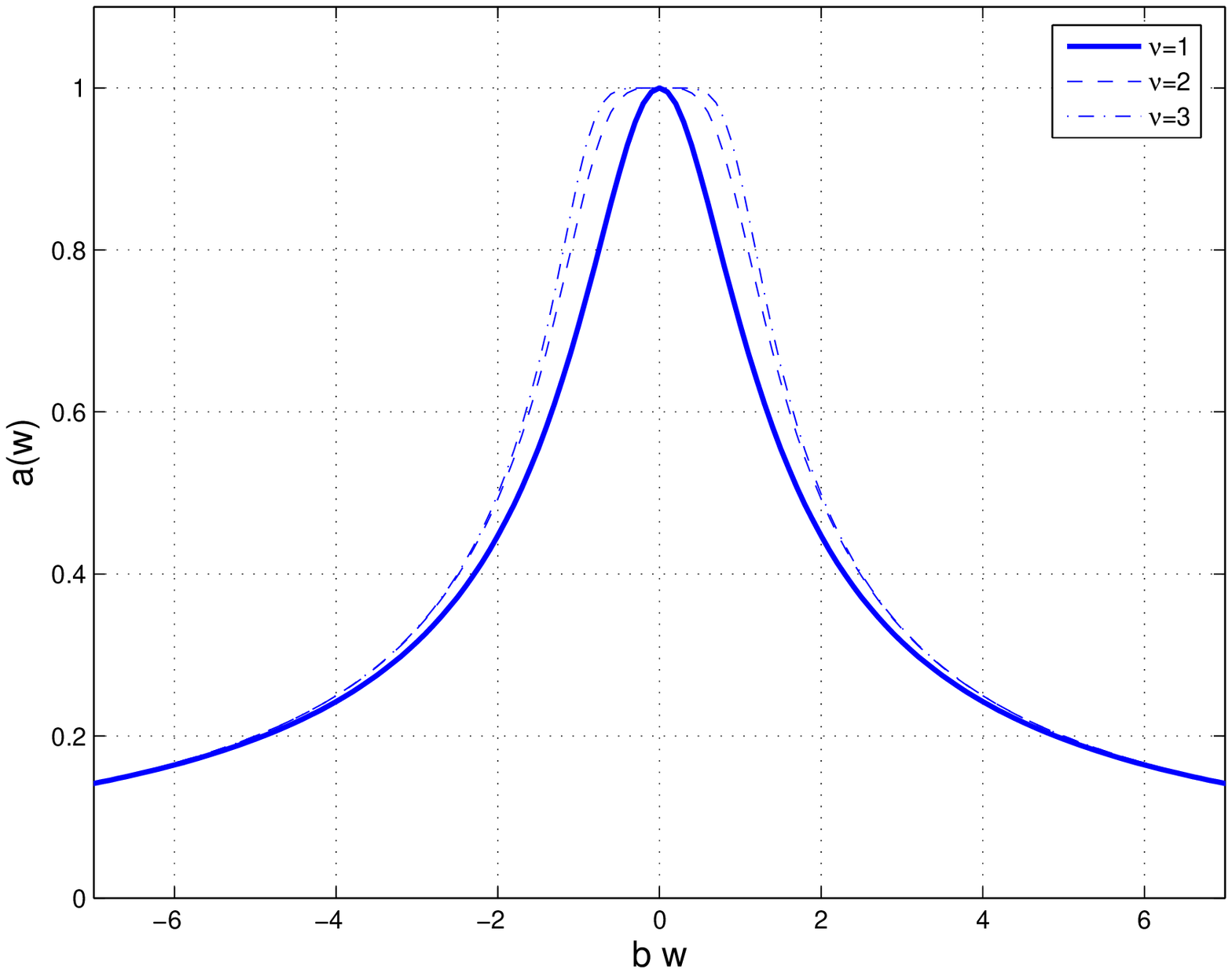}} &
      \hbox{\epsfxsize = 7.6 cm  \epsffile{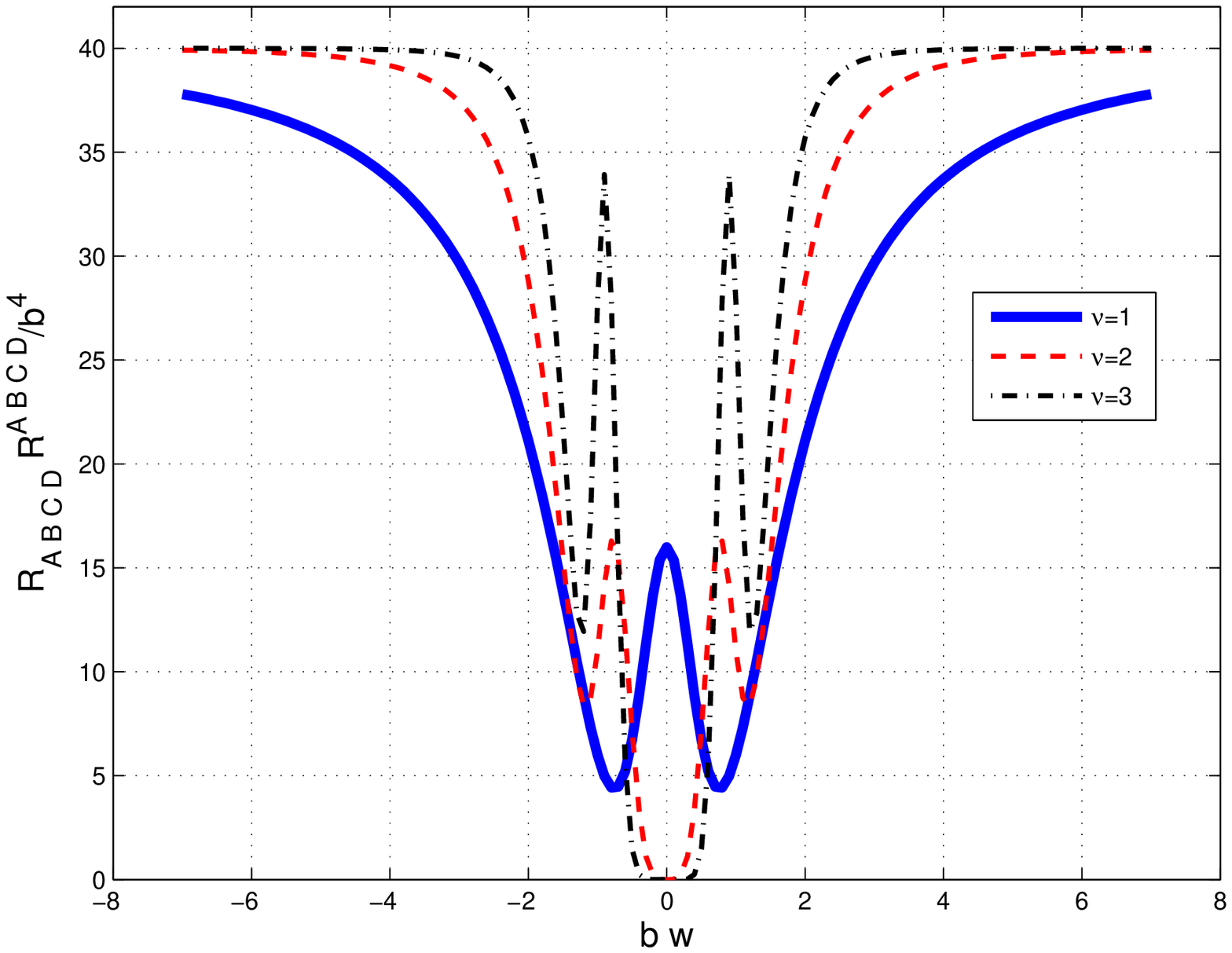}}\\
      \hline
\end{tabular}
\end{center}
\caption[a]{The profile of the warp factor (left plot) and the Riemann
invariant (right plot) for different values of $\nu$. For the other two curvature 
invariants reported in Eqs. (\ref{Rsqex}) and (\ref{Riccisqex}), the plots are qualitatively similar since they tend to a constant value for $|bw|\to \infty$ and they never get singular for finite $w$.}
\label{F4}
\end{figure}

  As already mentioned the geometry is ${\mathrm AdS_{5}}$
for $|bw|\to \infty$. This aspect can be clearly appreciated from Fig. \ref{F4}
(left plot) where the warp factor is illustrated for different values of $\nu$: 
for $|bw|\to \infty$, $a(w) \simeq 1/|bw|$. 
The curvature invariants, in the same limit, reach a constant value. 
In Fig. \ref{F4} (right plot) the Riemann invariant is illustrated. The 
other curvature invariants are qualitatively similar.
As a consequence of the features of the geometry the four-dimensional 
Planck mass is finite since it is simply given by
\begin{equation}
M^2_{P} \simeq M_{5}^3 \int_{-\infty}^{\infty} dw a^3(w)=  2 b M_{5}^3 
\frac{\Gamma\bigl(1 + \frac{1}{2\nu}\bigr) \Gamma\bigl(\frac{1}{\nu}\bigr)}{
\Gamma\bigl(\frac{3}{2\nu}\bigr)},\qquad \nu \geq 1
\label{pl}
\end{equation}
where the second equality follows by performing explicitly the integral 
when $a(w)$ is given by  Eq. (\ref{aw}) and when, as assumed throughout,
$\nu \geq 1$. Since the four-dimensional Planck mass is finite, the tensor fluctuations 
of the geometry are localized on the profile both for even and odd $\nu$. This occurrence is common 
also to the case when the defect is modeled by a 3-brane \cite{f5d}.
Less obvious is the fact that the scalar fluctuations of the sources are 
not localized. These results stem
from the analysis of the zero modes of the configurations 
defined by Eqs. (\ref{aw}), (\ref{phi}) and (\ref{Uphi}) and will now be 
swiftly addressed.  To discuss this 
problem one can then adopt the formalism developed in \cite{f4}.
In five dimensions the perturbed geometry 
leads to $15$ independent degrees of freedom which can be 
classified according to the way they transform under 
four-dimensional Poincar\'e transformations. 
To the fluctuations 
of the geometry one has also to add the fluctuation of the $\phi$, i.e.
the fluctuation of the profile of the defect. Therefore, the fluctuations of the 
geometry and of the scalar profile can be written as 
\begin{equation}
 g_{AB}(x^{\mu}, w) = \overline{g}_{AB}(w) + \delta g_{AB}(x^{\mu},w),
\qquad \phi(x^{\mu}, w) = \phi(w) + \chi(x^{\mu},w).
\label{chi}
\end{equation}
where 
\begin{equation}
\delta g_{A B}= a^2(w) \biggl(\matrix{2 h_{\mu\nu} 
+(\partial_{\mu} f_{\nu} +\partial_{\nu} f_{\mu}) 
+ 2\eta_{\mu \nu} \psi
+ 2 \partial_{\mu}\partial_{\nu} E
& D_{\mu} + \partial_{\mu} C &\cr
D_{\mu} + \partial_{\mu} C  & 2 \xi &\cr} \biggr).
\label{pert}
\end{equation}
On top of $h_{\mu\nu}$ which is divergence-less and trace-less (i.e. 
$\partial_{\mu}h^{\mu}_{\nu}$=0, $h_{\mu}^{\mu}=0$) there are four scalars
(i.e. $E$, $\psi$, $\xi$ and $C$) and two divergence-less vectors 
($D_{\mu}$ and $f_{\mu}$).  

The analysis can be conducted in gauge-invariant terms 
without assuming any specific form of the background geometry.
Neglecting the vector modes of the geometry \footnote{The vector modes 
are not localized since their corresponding zero mode is not normalizable as 
it follows from the evolution equations of $D_{\mu}$; the other vector, i.e. $f_{\mu}$ can 
be gauged away by using the freedom of fixing the coordinate system.}
the relevant zero modes 
are the ones associated with the graviton and with the scalar 
fluctuations. The decoupled evolution equation of the tensor modes 
can be written as \cite{f4}
\begin{equation}
\mu_{\mu\nu}'' - \partial_{\alpha}\partial^{\alpha} \mu_{\mu\nu} 
- \frac{(a^{3/2})''}{a^{3/2}} \mu_{\mu\nu} =0.
\label{mu}
\end{equation}
where $\mu_{\mu\nu} =a^{3/2} h_{\mu\nu}$ is the canonical 
normal mode of the of the action (\ref{action}) perturbed to second order 
in the amplitude of tensor fluctuations \cite{f4}.
The lowest mass eigenstate of Eq. (\ref{mu}) 
is  $\mu(w) = {\mu_0} a^{3/2}(w)$. Hence, the normalization 
condition of the tensor zero mode implies 
\begin{equation}
|\mu_0|^2 \int_{-\infty}^{\infty} a^3 ~dw = 2 |\mu_0|^2\int_{0}^{\infty}a^3(w)
~ dw=1.
\label{grav}
\end{equation}
The integral appearing in Eq. (\ref{grav}) 
is always convergent if, as assumed throughout 
the paper, $\nu$ is a positive integer. Therefore, the graviton 
zero mode is always localized on the configurations 
discussed here.
The scalar normal mode of the 
action is a linear combination $\chi$ (defined in Eq. (\ref{chi})) and 
and of $\psi$ (defined in Eq. (\ref{pert}). The canonical variable is then \cite{f4} 
${\cal G} = a^{3/2} \psi - z \chi$ and it obeys the equation
\begin{equation}
 {\cal G}'' - \partial_{\alpha}\partial^{\alpha}{\cal G}
- \frac{z''}{z} {\cal G} =0,\qquad z(w) = \frac{a^{3/2} \phi'}{{\cal F}}.
\label{G}
\end{equation}
The lowest mass eigenstate of Eq. (\ref{G}) is 
given by ${\cal G}(w)= {\cal G}_0 z(w)$ which is normalizable iff
\begin{equation}
 \int_{-\infty}^{\infty} z^2(w) ~dw = \frac{3}{\kappa}(2\nu -1) \int_{-\infty}^{+\infty}
  \frac{ d w}{(b w)^{2\nu} [ 1+ ( b w)^{2\nu}]^{\frac{3}{2\nu}}},\qquad \nu \geq 1
\label{INT}
\end{equation}
where the right hand side follows from the definition off $z(w)$ and from the 
explicit form of the solution.
But the integrand in Eq. (\ref{INT}) 
is divergent for $|bw|\rightarrow 0$ as $| b w|^{-2\nu}$.
As $\nu$ increases, the divergence becomes always more severe.
 We then conclude that the scalar modes of the geometry are not localized on the defect.

In conclusion a new class of solutions of five-dimensional 
warped geometries has been presented and discussed. This 
class of solution contains, simultaneously, kink-like 
profiles and bell-like scalar profiles. The regular geometry of the 
configuration allows the localization of the tensor modes of the geometry. 
Neither the scalar nor the vector  modes are localized. The present findings
seem to suggest that not only domain walls but also 
gravitating non-topological defects if five dimensions may be used 
to localize gravitational interactions. Furthermore, it is intriguing that these 
two rather different physical situations may arise in the same class of solutions.

\end{document}